\documentclass[aps,prl,twocolumn,showpacs,psfig,superscriptaddress,longbibliography]{revtex4-2}
\usepackage{multirow}
\usepackage{amsfonts}
\usepackage{mathrsfs}
\usepackage{siunitx}
\usepackage{amsmath}
\usepackage{color}
\usepackage{natbib}
\usepackage{textcomp}
\usepackage{graphicx}
\usepackage{bm}
\usepackage{amssymb}
\usepackage{xspace}
\usepackage{epstopdf}
\usepackage{dcolumn}
\usepackage{longtable}
\usepackage{setspace}
\usepackage{gensymb}
\usepackage{multirow}
\usepackage[colorlinks=true, letterpaper=true, pdfstartview=FitV, linkcolor=blue, citecolor=blue, urlcolor=blue]{hyperref}
\usepackage{float}
\usepackage{bm}

\makeatletter
\newcommand{\Rmnum}[1]{\expandafter\@slowromancap\romannumeral #1@}
\makeatother
\begin{document}

\title{Emergent magnetic states due to stacking and strain in the van der Waals magnetic trilayer CrI$_3$}

\author{Zhen Zhang}
\affiliation{School of Physical Sciences, University of Chinese Academy of Sciences, Beijng 100049, China}

\author{Jing-Yang You}
\email{phyjyy@nus.edu.sg}
\affiliation{Department of Physics, National University of Singapore, 2 Science Drive 3, 117551, Singapore}

\author{Bo Gu}
\email{gubo@ucas.ac.cn}
\affiliation{Kavli Institute for Theoretical Sciences, and CAS Center for Excellence in Topological Quantum Computation, University of Chinese Academy of Sciences, Beijng 100190, China}
\affiliation{Physical Science Laboratory, Huairou National Comprehensive Science Center, Beijing 101400, China}

\author{Gang Su}
\email{gsu@ucas.ac.cn}
\affiliation{School of Physical Sciences, University of Chinese Academy of Sciences, Beijng 100049, China}
\affiliation{Kavli Institute for Theoretical Sciences, and CAS Center for Excellence in Topological Quantum Computation, University of Chinese Academy of Sciences, Beijng 100190, China}
\affiliation{Physical Science Laboratory, Huairou National Comprehensive Science Center, Beijing 101400, China}

\begin{abstract}
Recently, three different magnetic states were observed experimentally in trilayer CrI$_3$ under pressure, including ferromagnetic (FM)-$\uparrow\uparrow\uparrow$, FM-$\downarrow\uparrow\downarrow$and FM-$\uparrow\uparrow\downarrow$. To reveal the nature of the observed three magnetic states, we studied the magnetic properties of four possible stacking structures in trilayer CrI$_3$: \Rmnum{1} (rhombohedral), \Rmnum{2} (monoclinic), \Rmnum{3} (hexagonal) and \Rmnum{4} (triclinic). We find that all four stacking structures possess the FM-$\uparrow\uparrow\uparrow$ ground state. After applying a few strains, the FM-$\downarrow\uparrow\downarrow$ becomes the ground state in \Rmnum{2} and \Rmnum{3} structures, and the FM-$\uparrow\uparrow\downarrow$ is preferred in \Rmnum{4} structure, while the FM-$\uparrow\uparrow\uparrow$ persists in \Rmnum{1} structure. Our results unveil that the three magnetic states observed in trilayer CrI$_3$ may correspond to different stacking structures with small tensile strains, which can well interpret the experimentally obtained pressure dependent interlayer coupling and Curie temperature. Our present study paves a way to design the magnetic multilayers with required magnetic states by tuning stacking and strain.
\end{abstract}
\maketitle


\section{\uppercase\expandafter{\romannumeral1}. Introduction}
After the recent discovery of ferromagnetic order in atomically thin layer CrI$_3$ \cite{Huang2017} and Cr$_2$Ge$_2$Te$_6$ \cite{Gong2017}, two-dimensional (2D) van der Waals (vdW) ferromagnetic semiconductors have attracted much attention due to their exotic properties and  potential applications in spintronics\cite{Gibertini2019,You2019,You2019b,Ningrum2020,You2021a}. According to Mermin-Wagner theorem \cite{Mermin1966}, the large magnetic anisotropy is required to stabilize the long-range ferromagnetism in 2D materials \cite{You2020,Deng2018,Bonilla2018,OHara2018}.

The magnetism in 2D materials can be sensitively controlled by external perturbations, such as electric field \cite{Weisheit2007,Deng2018,Zhang2020,You2021}, strain \cite{Mukherjee2019,Dong2019} and stacking \cite{Chen2019}. The stacking engineering is very promising, because there are many possible compositions to construct heterostructures with different 2D materials to enhance the magnetic properties\cite{Zhong2017,Mogi2018,Wang2020,Dong2020}, and to produce novel physical phenomena, such as the quantum anomalous Hall effect \cite{Deng2020,Zhao2020}, and axion insulators \cite{Liu2020}.

The 2D CrI$_3$, as an Ising-type ferromagnets with the Curie temperature of 45K in monolayer, has become a highlighted research hotspot.\cite{Song2018,Wang2018a,Klein2018,Kim2018,Sivadas2018,Jiang2019,Jang2019,Sun2019,Guo2019,Leon2020,Lei2021,Tian2021,Cantos-Prieto2021}. For a few layers of CrI$_3$, it can be used as a spin-filter tunnel barrier possessing a giant tunneling magnetoresistance \cite{Song2018,Wang2018a,Klein2018,Kim2018}, and the magnetic interactions between adjacent layers can be tuned by electric gating, electrostatic doping and pressure\cite{Jiang2018,Jiang2018a,Huang2018a,Li2019,Song2019}. Most of the theoretical studies on few-layers CrI$_3$ focused on bilayer \cite{Sivadas2018,Jiang2019,Jang2019,Leon2020}, and they show that rhombohedral stacking favors ferromagnetic (FM) interlayer interaction, while monoclinic stacking is beneficial to antiferromagnetic (AFM) interlayer interaction, which is in good agreement with the experimental observations\cite{Sun2019}. \textcolor{black}{Bulk CrI$_3$ is reported to be a ferromagnetic semiconductor with a band gap of about 1.2 eV \cite{Dillon1965}, which maintains the rhombohedral stacking sequence with $R$$\overline{3}$ space group symmetry with the temperature lower than $\sim$220 K, and transforms to the monoclinic stacking with $C$2/$m$ space group at higher temperature \cite{McGuire2015}.}


Recently, researchers have turned their attention to trilayer CrI$_3$. In 2019, Peng et al observed the rhombohedral stacking order with FM interlayer interaction (FM-$\uparrow\uparrow\uparrow$) at 10K in non-encapsulated trilayer CrI$_3$ \cite{Guo2019}. However, Xu et al reported that the pristine exfoliated trilayer CrI$_3$ favors AFM interlayer interaction with spin antiparallel arrangement of every two adjacent layers (labeled as $\downarrow\uparrow\downarrow$), and under pressures, three magnetic states, i.e. FM-$\uparrow\uparrow\uparrow$, FM-$\downarrow\uparrow\downarrow$, and FM-$\uparrow\uparrow\downarrow$, can be obtained in trilayer CrI$_3$ \cite{Song2019}. It becomes important to uncover the nature of different magnetic states in trilayer CrI$_3$ observed in recent experiments.

In this paper, we systematically investigate the magnetic properties of the trilayer CrI$_3$ with four different stacking structures, including \Rmnum{1} (rhombohedral), \Rmnum{2} (monoclinic), \Rmnum{3} (hexagonal) and \Rmnum{4} (triclinic). Our results show that the magnetic states FM-$\uparrow\uparrow\uparrow$, FM-$\downarrow\uparrow\downarrow$, and FM-$\uparrow\uparrow\downarrow$ experimentally observed in trilayer CrI$_3$ may come from different stacking structures with small tensile strain in trilayer CrI$_3$. The correspondence between the stacking structure and interlayer magnetic coupling could provide us a basis to design spintronic devices with desirable magnetic properties by adjusting the stacking order and strains.

\section{\uppercase\expandafter{\romannumeral2}. Computational Methods}
In our calculations, the projector augmented wave (PAW) method~\cite{Bloechl1994} based on the density functional theory (DFT) as implemented in the Vienna ab initio simulation package (VASP) \cite{Kresse1993,Kresse1996} is employed to carry out the first-principles calculations. The generalized gradient approximation (GGA) in the form proposed by Perdew, Burke, and Ernzerhof (PBE) \cite{Perdew1996} is used to describe the electron exchange-correlation functional. To simulate the quasi-2D materials, a sufficiently large vacuum of 15 {\AA} along the z direction is built. The zero damping DFT-D3 method is adopted to describe the interlayer van der Waals interaction. The plane-wave cutoff energy is set to be 450 eV. Structural optimization including lattice constant and atomic positions is done with the conjugate gradient (CG) scheme until the maximum force acting on each atoms is less than 0.01 eV/\AA~and the total energy is converged to 10$^{-5}$ eV. The $9\times9\times1$ $k$-point mesh grid within the first Brillouin zone ~\cite{Monkhorst1976} is used for structure optimization and self-consistent calculation.  Self-consistent-field DFT calculations incorporating the spin-orbit coupling (SOC) are used to obtain the energies accurately. Considering the on-site Coulomb interaction of Cr 3$d$ orbitals, $U$ = 3 eV is used in all calculations, and the effect of different $U$ values is also tested to show the robustness of our results.

\section{\uppercase\expandafter{\romannumeral3}. Results}

\subsection{A. Four stacking orders in trilayer CrI$_3$}
\begin{figure*}[!!!hbt]
  \centering
  \includegraphics[scale=0.60,angle=0]{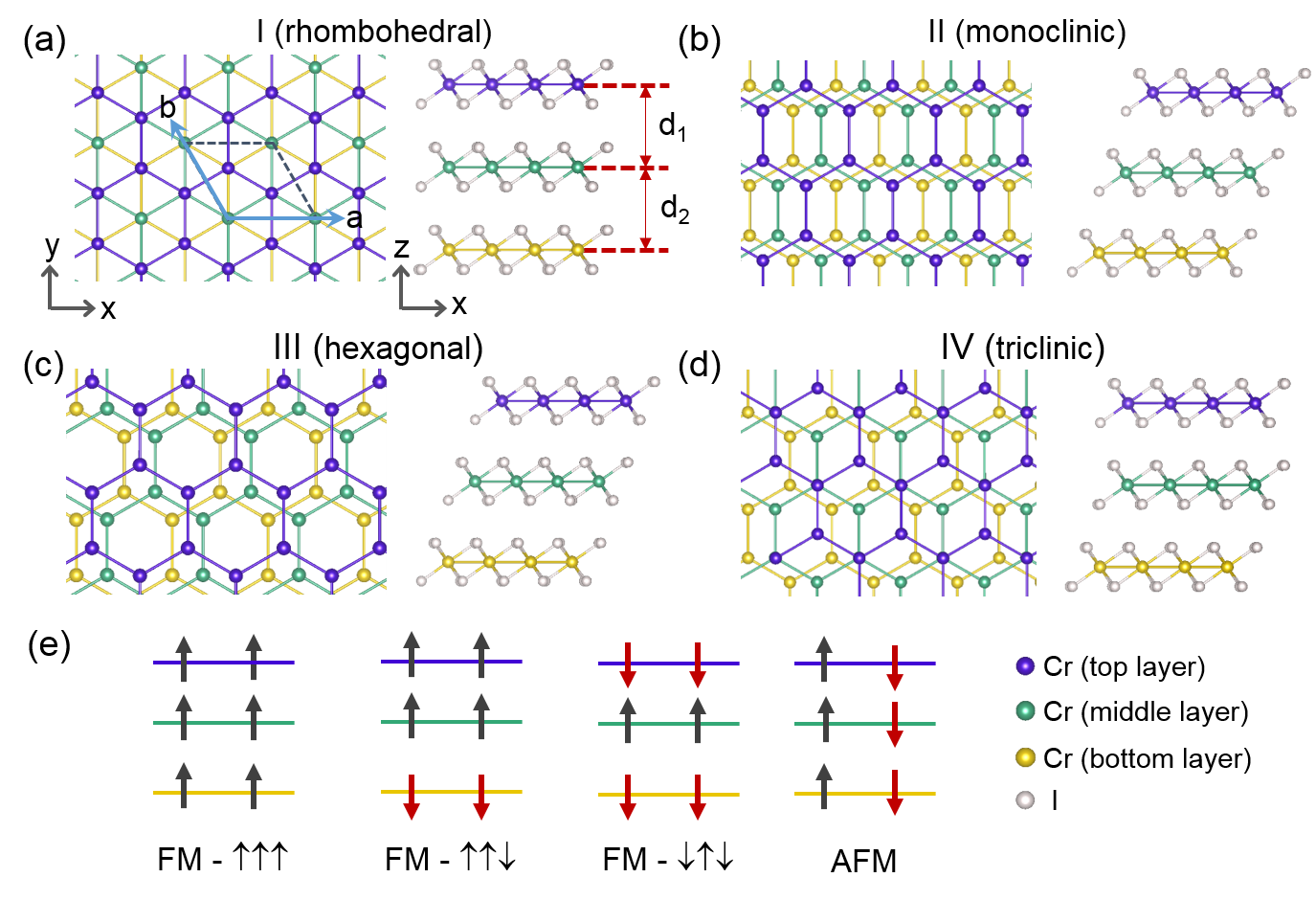}\\
  \caption{The top and side views of (a) rhombohedral, (b) monoclinic, (c) hexagonal and (d) triclinic stacking structures in trilayer CrI$_3$. The yellow, green, and purple balls are the bottom, middle, and top Cr layers, respectively. (e) Schematic plot of four different spin configurations. FM and AFM denote the ferromagnetic and antiferromagnetic intralayer exchange couplings, respectively, and $\uparrow\uparrow\uparrow$, $\uparrow\uparrow\downarrow$ and $\downarrow\uparrow\downarrow$ represent the three possible interlayer spin orientations.}\label{fig1}
\end{figure*}

The primitive cell of monolayer CrI$_3$ consists of two magnetic Cr atoms and six I atoms with each Cr atom surrounded by six I atoms forming a distorted octahedron. The basic vectors $\textit{\textbf{a}}$ and $\textit{\textbf{b}}$ are along the zigzag direction of the honeycomb lattice composed of Cr atoms and the y-axis corresponds to the armchair direction as shown in Fig. \ref{fig1}(a). Here, we consider four different stacking structures in trilayer CrI$_3$: \Rmnum{1} (rhombohedral), which is the low-temperature phase of bulk CrI$_3$ with the upper layer of two adjacent layers always translating $\sqrt{3}/3a$ in the armchair direction (y-axis) relative to the lower layer; \Rmnum{2} (monoclinic), the high-temperature phase of bulk CrI$_3$ with the upper layer of two adjacent layers always moving 1/3a in the zigzag ($\textit{\textbf{a}}$) direction relative to the lower layer; \Rmnum{3} (hexagonal), which is the primitive cell of the bulk CrI$_3$ with \textit{P}3$_1$12 space group, where the middle and top layers move 1/3a along the zigzag direction $\textit{\textbf{a}}$ and $\textit{\textbf{b}}$ directions relative to the lower layer, respectively; \Rmnum{4} (triclinic), which is constructed with the middle and top layer moving 1/3a and $\sqrt{3}/3a$ along the zigzag ($\textit{\textbf{a}}$) and armchair (y) directions relative to the lower layer, respectively, as shown in Figs. \ref{fig1}(a)-(d).

\textcolor{black}{To study the possible stacking structures of trilayer CrI$_3$, the interlayer exchange couplings are essential, which should be included in the model. Because the interlayer Cr-Cr distances (d$_1^{\prime}$, d$_2^{\prime}$ and d$_1^{\prime\prime}$) are comparable with the intralyer second and third neighboring Cr-Cr distances (d$_2$, d$_3$), as noted in Table S3 in Supplemental Materials, it is necessary to consider the intralyer second and third nearest-neighboring exchange couplings (J$_2$ and J$_3$). In order to explore the relation between magnetic states and stacking structures of trilayer CrI$_3$, for each stacking structure we have studied four magnetic states FM-$\uparrow\uparrow\uparrow$, FM-$\uparrow\uparrow\downarrow$, FM-$\downarrow\uparrow\downarrow$ and AFM in our calculations,} where FM (AFM) and $\uparrow\uparrow\uparrow$ ($\uparrow\uparrow\downarrow$, $\downarrow\uparrow\downarrow$) represent the intralayer and interlayer exchange couplings, respectively, as shown in Fig. \ref{fig1}(e). By comparing the total energies of different spin configurations for every stacking structure, FM-$\uparrow\uparrow\uparrow$ state is found to be the magnetic ground state for the above four stacking structures, and the experimental and calculated  structural parameters are listed in Table \ref{tab:structure}. Among the four stacking structures, \Rmnum{1} (rhombohedral) stacking structure possesses the lowest energy, and we select its lattice constant 6.962 \AA~as the initial lattice constant to investigate the effect of in-plane strain.

\textcolor{black}{For the mixture of rhombohedral and hexagonal stackings, it contains the relative movement between adjacent layers in both armchair and zigzag directions, which is identical to the \Rmnum{4} (triclinic) stacking order. For the mixture of hexagonal and monoclinic stackings, it contains the relative movement between adjacent layers in the zigzag direction, leading to \Rmnum{2} (monoclinic) and \Rmnum{3} (hexagonal) stacking structures.}

\begin{table}[t]
\begin{center}
\caption{Structural parameters for bulk and trilayer CrI$_3$. Lattice constant ($a$), interlayer distance ($d_1$ and $d_2$) reported in the experiment for bulk CrI$_3$ with rhombohedral and monoclinic stacking orders \cite{McGuire2015} and our DFT results for trilayer CrI$_3$ with I (rhombohedral), II(monoclinic), III (hexagonal) and IV (triclinic) stacking orders. $d_1$ and $d_2$ refer to the distance between the middle layer and its upper and lower layers, respectively, as shown in Fig. \ref{fig1}(a).}\label{tab:structure}
\centering
\begin{spacing}{1.3}
\setlength{\tabcolsep}{1.5mm}{
\begin{tabular}{m{2.0cm}<{\centering}|m{2.4cm}|m{1.0cm}<{\centering}m{1.0cm}<{\centering}m{1.0cm}<{\centering}}
\hline
\multicolumn{2}{c|}{CrI$_3$}                                 &  $a$ ({\AA})  & $d_1$ ({\AA})  & $d_2$ ({\AA}) \\
\hline
     Bulk                                    & rhombohedral       &  6.867       &  6.602        & 6.602 \\
(Exp. in\cite{McGuire2015})                                       & monoclinic         &  6.866       &  6.623        & 6.623 \\
\hline
                                       & I(rhombohedral)    &  6.962      &  6.706         & 6.706 \\
Trilayer                               & II (monoclinic)    &  6.957      &  6.762         & 6.762 \\
(DFT)                                  & III (hexagonal)    &  6.960      &  6.755         & 6.755 \\
                                       & IV (triclinic)     &  6.960      &  6.681         & 6.720 \\
\hline
\end{tabular}}
\end{spacing}
\end{center}
\end{table}

\subsection{B. Magnetic ground state}
For simplicity, we apply the in-plane biaxial strain to investigate the strain effect on magnetic properties in trilayer CrI$_3$ for four different stacking structures. The in-plane biaxial strain is defined as $\varepsilon = (a-a_0)/a_0$, where $a_0$ and $a$ are lattice parameters without and with strain, respectively. For the trilayer CrI$_3$ with \Rmnum{1} (rhombohedral) stacking structure, the interlayer distance ($d$) decreases from 7.05 to 6.38 {\AA} with the in-plane biaxial strain changing from -10\% to 10\% as shown in Fig. S1. With the increase of tensile strain, the FM-$\uparrow\uparrow\uparrow$ spin configuration is always the magnetic ground state, and the energy difference between other spin configurations and FM-$\uparrow\uparrow\uparrow$ is increased, indicating a more stable FM-$\uparrow\uparrow\uparrow$ magnetic state with the decrease of interlayer distance. While with the increase of compressive strain, a magnetic phase transition from FM-$\uparrow\uparrow\uparrow$ to AFM will occur when the compressive strain is larger than 6\%. The similar magnetic phase transition from intralayer FM to AFM with the increase of compressive strain also occurred in monolayer CrI$_3$ \cite{Webster2018}. For trilayer CrI$_3$ with \Rmnum{2} (monoclinic) and \Rmnum{3} (hexagonal) stacking structures, a tiny tensile strain will cause a magnetic phase transition from FM-$\uparrow\uparrow\uparrow$ to FM-$\downarrow\uparrow\downarrow$, and the reduced interlayer distance can stabilize the FM-$\downarrow\uparrow\downarrow$ state. The change of magnetic phases under compressive strain for \Rmnum{2} and \Rmnum{3} structures is similar to that for the \Rmnum{1} (rhombohedral) stacking structure. \textcolor{black}{The above \Rmnum{1}, \Rmnum{2}, \Rmnum{3} stacking structures cannot lead to the FM-$\uparrow\uparrow\downarrow$ magnetic phase, which was observed in the experiment. To understand the stacking structure of the magnetic state FM-$\uparrow\uparrow\downarrow$, which involves the both ferromagnetic and antiferromagnetic interlayer couplings in trilayer CrI$_3$, we have considered the \Rmnum{4} (triclinic) stacking structure, which combines the structures of \Rmnum{1} and \Rmnum{2} (or \Rmnum{3}). Therefore, three magnetic states FM-$\uparrow\uparrow\uparrow$, FM-$\downarrow\uparrow\downarrow$ and FM-$\uparrow\uparrow\downarrow$ observed experimentally in the trilayer CrI$_3$ \cite{Song2019} are all obtained in our calculations.}


\begin{figure}[!!!hbt]
  \centering
  \includegraphics[scale=0.58,angle=0]{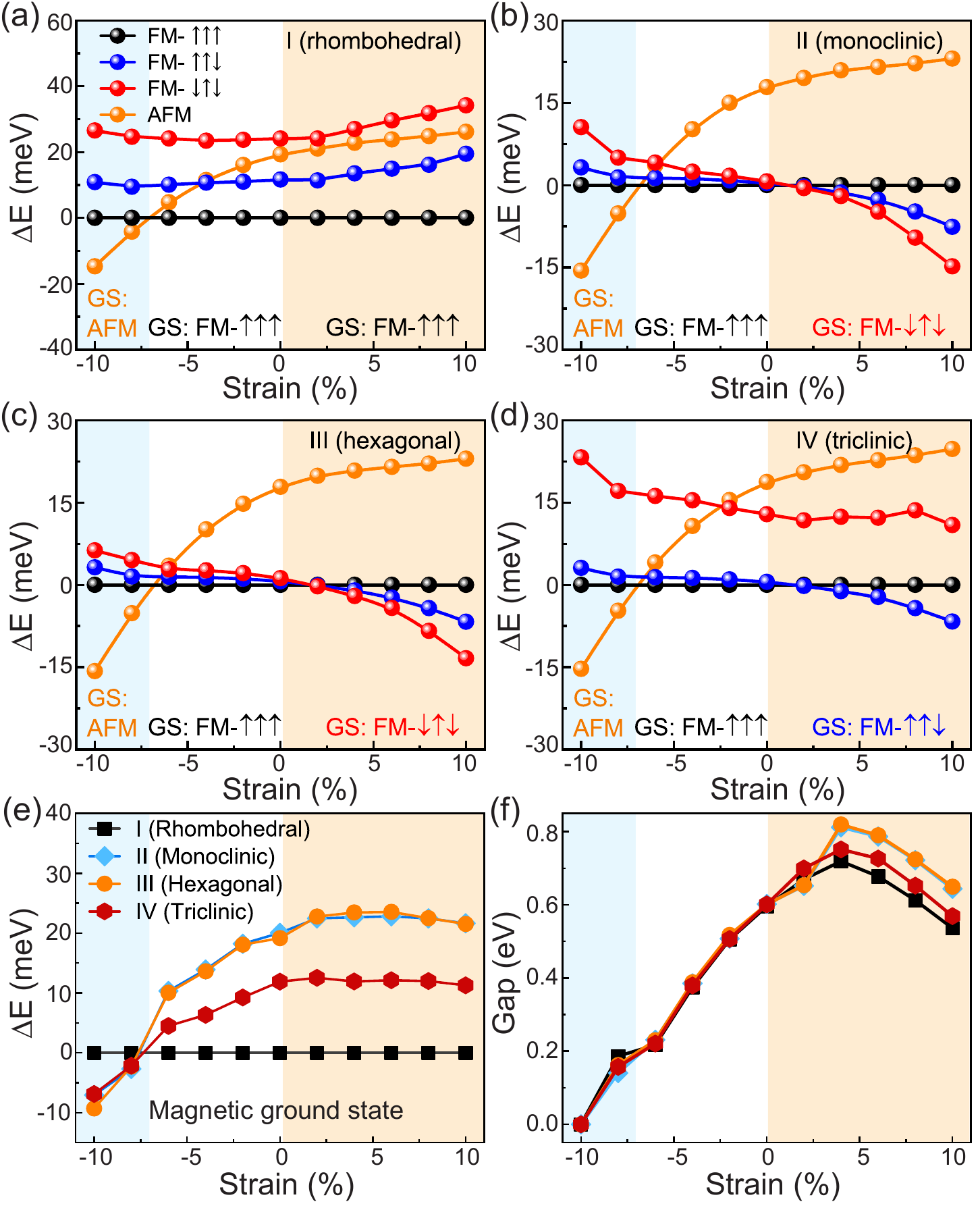}\\
  \caption{The relative total energy as a function of strain for (a) rhombohedral, (b) monoclinic, (c) hexagonal and (d) triclinic stacking structures in trilayer CrI$_3$. For every stacking order, four different spin configurations FM-$\uparrow\uparrow\uparrow$, FM-$\downarrow\uparrow\downarrow$, FM-$\uparrow\uparrow\downarrow$, and AFM are considered. The energy difference is defined as $\Delta$$E$ = $E$ - $E_{0}$, where $E_{0}$ is the energy of FM-$\uparrow\uparrow\uparrow$ state at corresponding strain. (e) The relative total energy of \Rmnum{2}, \Rmnum{3} and \Rmnum{4} stacking structures. $\Delta$$E$ is defined as $\Delta$$E$ = $E$ - $E_{\Rmnum{1}}$, where $E_{\Rmnum{1}}$ is the energy of \Rmnum{1} stacking structure under different strains. (f) Strain-dependent band gaps for four stacking structures in trilayer CrI$_3$.}\label{fig2}
\end{figure}

To study the relative stability of four stacking structures, we calculate the total energy of four stacking orders under different strains as shown in Fig. \ref{fig2} (e), and find that in the range of compressive strain -6\% to tensile strain 10\%, the energy of the \Rmnum{1} (rhombohedral) stacking structure is always the lowest, and \Rmnum{2} (monoclinic) and \Rmnum{3} (hexagonal) structures have almost the same energy and is the highest. The total energy of the IV (triclinic) stacking structure is between that of \Rmnum{1} (rhombohedral) and \Rmnum{2} (monoclinic) structures, and it is interesting to note that the IV structure can be regarded as the combination of structures \Rmnum{1} and \Rmnum{2}.
As shown in Fig. \ref{fig2} (f), the electronic band gap shows a strong dependence on the magnetic states in trilayer CrI$_3$. In the range of 0$\sim$-10\% compressive strain, four stacking structures \Rmnum{1} to \Rmnum{4} possess the same magnetic ground states, and their band gaps are basically the same, and they all undergo a transition from semiconductor to metal with compressive tensile about -10\%. In the range of 0$\sim$10\% tensile strain, the band gaps for stacking structures \Rmnum{1}-\Rmnum{4} become different, and at the same time their magnetic ground states become different.

\subsection{C. Theoretical model analysis}
\begin{figure}[!!!hbt]
  \centering
  \includegraphics[scale=0.40,angle=0]{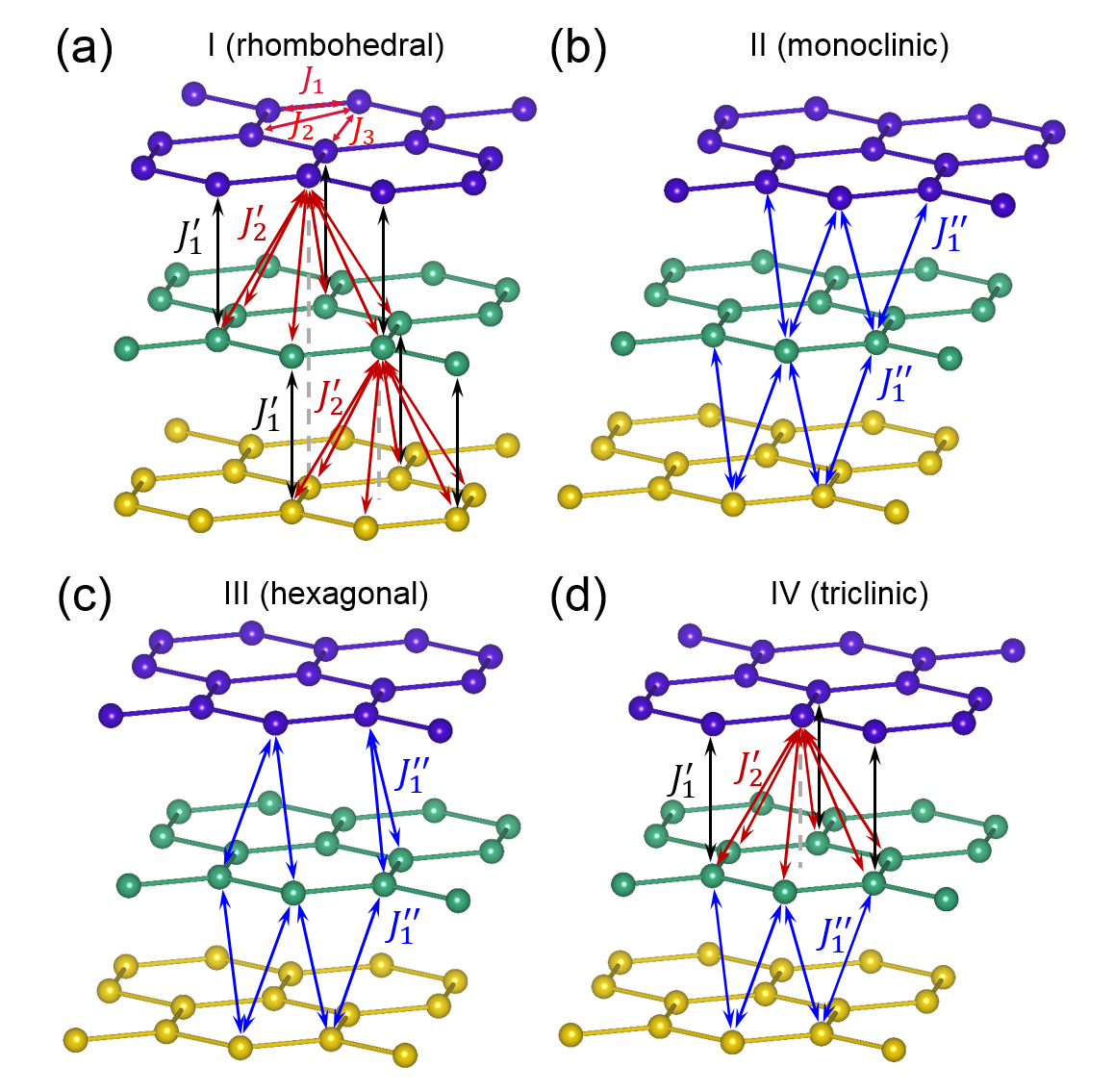}\\
  \caption{Interlayer interactions in (a) \Rmnum{1} (rhombohedral) with the first (J$_1^{\prime}$) and second (J$_2^{\prime}$) nearest-neighboring exchange couplings labeled, (b) \Rmnum{2} (monoclinic) with the nearest-neighbor (J$_1^{\prime\prime}$) labeled, (c) \Rmnum{3} (hexagonal) with the nearest-neighbor (J$_1^{\prime\prime}$) labeled and (d) \Rmnum{4} (triclinic) with J$_1^{\prime}$, J$_2^{\prime}$ and J$_1^{\prime\prime}$ co-existing, respectively. The intralayer interactions including first (J$_1$), second (J$_2$) and third (J$_3$) nearest-neighboring exchange couplings between two Cr atoms are marked in (a).}\label{fig3}
\end{figure}

\textcolor{black}{To better understand the relation between the magnetic states and the stacking structures in trilayer CrI$_3$, we consider a Hamiltonian including intralayer and interlayer interactions $H=H_{intra}+H_{inter}$. The intralayer term $H_{intra}$ is written as}
\textcolor{black}{\begin{equation}
\begin{split}
\label{H}
&H_{intra} = H_0+H_{MAE},\\
H_0 = -\sum_{\langle i,j\rangle}J{_1}\boldsymbol{S_i}\cdot\boldsymbol{S_j}&-\sum_{\langle\langle i,j\rangle\rangle}J{_2}\boldsymbol{S_i}\cdot\boldsymbol{S_j}-\sum_{\langle\langle\langle i,j\rangle\rangle\rangle}J{_3}\boldsymbol{S_i}\cdot\boldsymbol{S_j},\\
\end{split}
\end{equation}
where $H_0$ is the isotropic Heisenberg model with $J_1$, $J_2$ and $J_3$ the intralayer the first-, second-, and third-nearest-neighboring exchange interactions, respectively, and $H_{MAE}$ is the magnetic anisotropy including the Kitaev-like exchange anisotropy and the single-ion anisotropy  \cite{Xiang2013, Xu2018}. The calculation details are provided in Supplemental Materials.}
For the two adjacent layers sliding $\sqrt{3}/3a$ along the armchair direction, the interlayer interaction $H_{inter}$ could be written as
\begin{equation}
\label{Hinter}
H_{inter}=-\sum_{\langle i,i^\prime\rangle}J{_1}{^\prime}\boldsymbol{S}_i\cdot\boldsymbol{S}_{i^\prime}-\sum_{\langle\langle i,i^\prime\rangle\rangle}J{_2}{^\prime}\boldsymbol{S}_i\cdot\boldsymbol{S}_{i^\prime},
\end{equation}
where $J_1$$^\prime$ and $J_2$$^\prime$ are the interlayer first- and second-nearest-neighbor exchange interactions, respectively, and $i$ and $i^\prime$ represent Cr atoms in the adjacent layers. While for the two adjacent layers sliding 1/3a along the zigzag direction, the interlayer interaction $H_{inter}$ is written as
\begin{equation}
\begin{split}
\label{Hinter1}
H_{inter}=-\sum_{\langle i,i^\prime\rangle}J{_1}{^{\prime\prime}}\boldsymbol{S}_i\cdot\boldsymbol{S}_{i^\prime},
\end{split}
\end{equation}
where $J{_1}{^{\prime\prime}}$ represents the interlayer nearest exchange interactions. The exchange couplings existing in stacking structures \Rmnum{1} to \Rmnum{4} are marked in Fig. \ref{fig3}. \textcolor{black}{Because the interlayer exchange couplings are included in the model for the stacking structures of trilayer CrI$_3$, the intralyer second and third nearest-neighboring exchange couplings J$_2$ and J$_3$ are also included, as discussed in subsection III.A.} The values of J$_1$, J$_2$, J$_3$, J$_1$$^\prime$, J$_2$$^\prime$ and J$_1$$^{\prime\prime}$ can be extracted from DFT results by calculating the energies of several different spin configurations as shown in Figs. S2-S4 in Supplemental Materials, and the results for stacking structures \Rmnum{1} to \Rmnum{4} under different strains are listed in Table \ref{tab:exchange}. By Eqs. (\ref{H}) - (\ref{Hinter1}), positive (negative) exchange integrals indicate a FM (AFM) interaction. When the applied strain is between -6\%$\sim$10\%, the calculated intralayer exchange couplings $J_1$ and $J_2$ are all positive values, which are much larger than that of $J_3$ which is negligible small. \textcolor{black}{Table \ref{tab:exchange} and Fig. \ref{fig2} show that J$_1$ is weakened and becomes a negative value when the compressive strain is larger than 6\%. The intralayer nearest-neighboring exchange coupling is determined by the competition between AFM direct exchange and FM superexchange \cite{Goodenough1958,Kanamori1959,Lado2017}. Decreasing Cr-Cr intralayer distance, the AFM direct exchange can be enhanced, and will dominate the intralayer exchange interaction.}


\begin{table*}[t]
\begin{center}
\caption{The magnetic ground state, intralayer and interlayer exchange couplings (meV) and magnetic anisotropy energy (meV/Cr) under different strains for four staking structures. J$_1$, J$_2$, J$_3$, J$_1^{\prime}$, J$_2^{\prime}$ and J$_1^{\prime\prime}$ are labeled in Fig. \ref{fig3}, and a positive value indicates ferromagnetic coupling, whereas a negative value indicates antiferromagnetic coupling. Positive value of \textcolor{black}{E$_{MAE}$} indicates the out-of-plane magnetization, otherwise the in-plane magnetization. The four stacking structures and different spin configurations are shown in Fig. \ref{fig1}. }\label{tab:exchange}
\centering
\begin{spacing}{1.3}
\setlength{\tabcolsep}{1.8mm}{
\begin{tabular}{lccccccccc}
\hline
~~ Trilayer CrI$_3$ & Strain & Ground state & J$_1$$\vert$S$\vert$$^2$ & J$_2$$\vert$S$\vert$$^2$ & J$_3$$\vert$S$\vert$$^2$ &J$_1^{\prime}$$\vert$S$\vert$$^2$ & J$_2^{\prime}$$\vert$S$\vert$$^2$ & J$_1^{\prime\prime}$$\vert$S$\vert$$^2$ & \textcolor{black}{E$_{MAE}$} \\
\hline
                 & -6\% & FM-$\uparrow\uparrow\uparrow$   & 2.09 & 2.26 & -0.89 & -0.43 & 0.57   & / &1.22\\
I (rhombohedral) & ~0\% & FM-$\uparrow\uparrow\uparrow$   & 9.14 & 1.51 & -0.19 & -0.57 & 0.67   & / &0.63\\
                 & ~6\% & FM-$\uparrow\uparrow\uparrow$   & 11.23& 1.12 & -0.05 & -0.46 & 0.82   & / &0.62\\
\hline
                 & -6\% & FM-$\uparrow\uparrow\uparrow$   & 2.06 & 2.18 & -0.70  & / & / & ~0.13 & 1.22\\
II(monoclinic)   & ~0\% & FM-$\uparrow\uparrow\uparrow$   & 9.21 & 1.34 & ~0.06  & / & / & ~0.01 & 0.61\\
                 & ~6\% & FM-$\downarrow\uparrow\downarrow$ & 11.18 & 1.02 & ~0.14  & / & / & -0.41 & 0.29 \\
\hline
                 & -6\% & FM-$\uparrow\uparrow\uparrow$   & 2.05 & 2.19 & -0.70  & / & / & 0.14  & 1.25 \\
III(hexagonal)   & ~0\% & FM-$\uparrow\uparrow\uparrow$   & 9.20 & 1.35 & ~0.07  & / & / & ~0.02 & 0.66\\
                 & ~6\% & FM-$\downarrow\uparrow\downarrow$ & 11.18 & 1.02 & ~0.13  & / & / & -0.38& 0.46\\
\hline
                 & -6\% & FM-$\uparrow\uparrow\uparrow$ & 2.06 & 2.18 & -0.69  & -0.50 & 0.59 & 0.13 & 1.23\\
IV(triclinic)    & ~0\% & FM-$\uparrow\uparrow\uparrow$ & 9.28 & 1.33 & ~0.06  & -0.63 & 0.70 & ~0.01 & 0.61\\
                 & ~6\% & FM-$\uparrow\uparrow\downarrow$ & 11.20 & 1.01 & ~0.13  & -0.53 & 0.83 & -0.39 &0.45 \\
\hline
\end{tabular}}
\end{spacing}
\end{center}
\end{table*}

\subsection{D. Comparison  with the experiment}
\begin{figure}[!!!hbt]
  \centering
  \includegraphics[scale=0.55,angle=0]{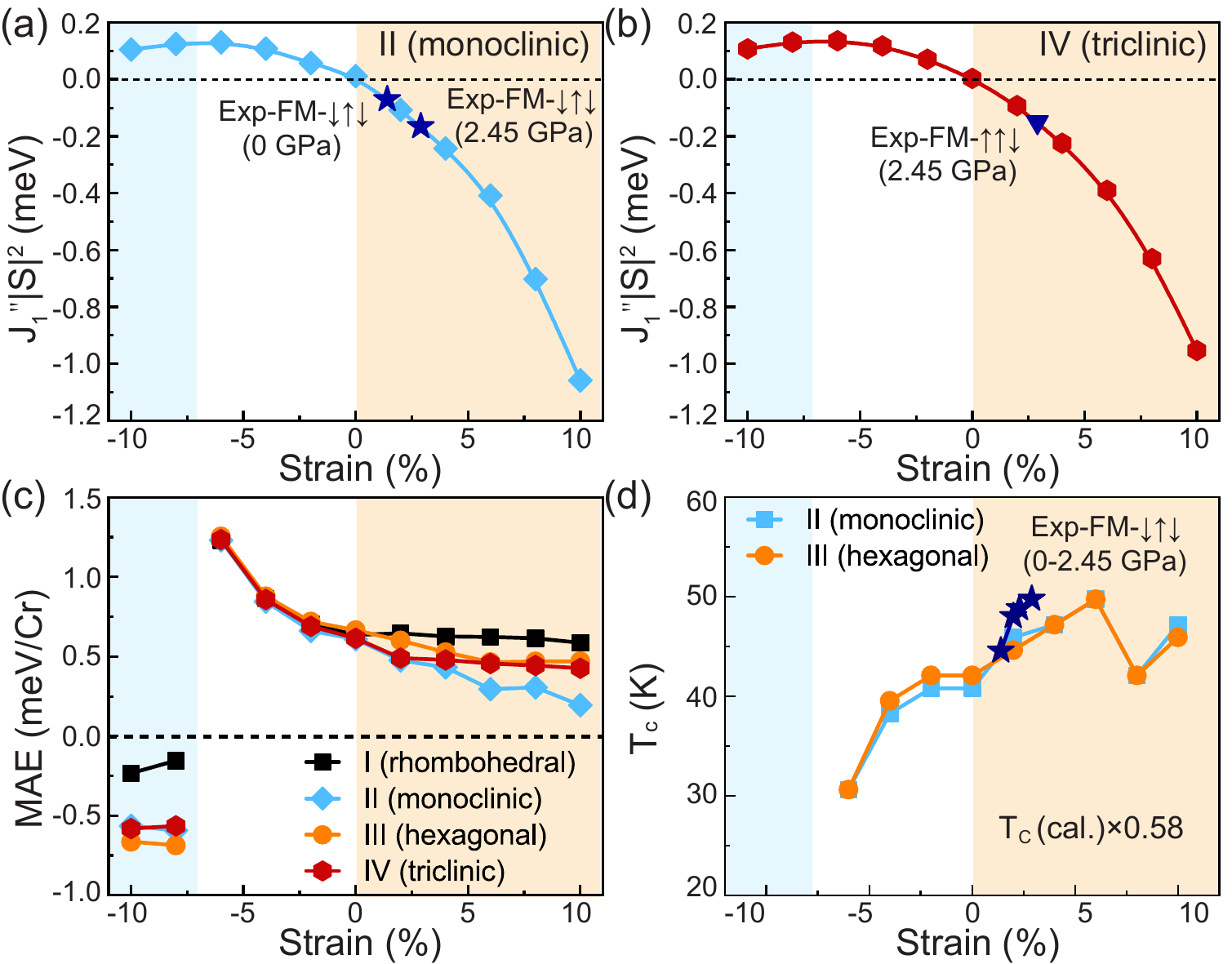}\\
  \caption{ Strain dependent interlayer exchange coupling J$_1^{\prime\prime}$$\vert$S$\vert$$^2$ for (a) monoclinic and (b) triclinic stacking orders. The navy blue points are simulated from the experimental results \cite{Song2019}. (c) Strain-dependent magnetic anisotropy energy (MAE) which is defined as $E_{\bm{m}\parallel\bm{x}}$-$E_{\bm{m}\parallel\bm{z}}$ for four different stacking structures in trilayer CrI$_3$ . A positive value of MAE indicates the out-of-plane magnetization, otherwise the in-plane magnetization. \textcolor{black}{(d) Strain-dependent Curie temperature (T$_C$) for \Rmnum{2} (monoclinic) and \Rmnum{3} (hexagonal) stacking structures in trilayer CrI$_3$ by Monte Carlo simulations, which has been rescaled by a factor of 0.58 to reproduce the experimental results}. The navy blue points in (d) are the experimental T$_C$ in trilayer CrI$_3$ with different applied pressures \cite{Song2019}.}\label{fig4}
\end{figure}

Three different magnetic phases FM-$\uparrow\uparrow\uparrow$, FM-$\downarrow\uparrow\downarrow$, FM-$\uparrow\uparrow\downarrow$ have been experimentally observed in the trilayer CrI$_3$ under the application of pressure \cite{Song2019}. For the measured FM-$\downarrow\uparrow\downarrow$ magnetic phase, which corresponds to \Rmnum{2} (monoclinic) and \Rmnum{3} (hexagonal) stacking structures with a tiny tensile strain based on our DFT results, the critical field ($B_C$) polarizing the FM-$\downarrow\uparrow\downarrow$ state to FM-$\uparrow\uparrow\uparrow$ state under different pressure is reported in the experiment \cite{Song2019}. Here, we adopt $4J_1^{\prime\prime}$$\vert$S$\vert$$^2$ $\approx$ -$g\mu{_B}SB{_C}$ to roughly estimate the strength of the interlayer exchange coupling $J_1^{\prime\prime}$ under different pressures. The experimental values of $B_C$ under 0 and 2.45 GPa are 1.6 and 3.7 T, respectively. Thus the corresponding interlayer exchange couplings $J_1^{\prime\prime}$$\vert$S$\vert$$^2$ could be obtained as -0.07 and -0.16 meV, respectively. The values of J$_1^{\prime\prime}$$\vert$S$\vert$$^2$obtained from the DFT results for \Rmnum{2} (monoclinic) stacking structure are plotted as a function of the applied strain as shown in Fig .\ref{fig4}(a). In this way, for the applied pressure in the experiment, the corresponding tensile strain can be estimated. As noted in Fig. \ref{fig4} (a), for the 0 GPa in the experiment, a small tensile strain about 1.4\% is obtained in our estimation. The strain may be due to the existence of the encapsulation of trilayer CrI$_3$ by graphene and h-BN in the experimental setup \cite{Song2019}. The estimated strain is increased to 2.9\% when the pressure is applied to 2.45 GPa in the experiment. For the measured FM-$\uparrow\uparrow\downarrow$ magnetic phase under 2.45 GPa pressure, which corresponds to \Rmnum{4} (triclinic) stacking structure with tensile stain based on our DFT results, $2J_1^{\prime\prime}$$\vert$S$\vert$$^2$ $\approx$ -$g\mu{_B}SB{_C}$ is used to estimate the value of $J_1^{\prime\prime}$$\vert$S$\vert$$^2$ and the result is marked in Fig. \ref{fig4}(b). Magnetic anisotropy energy (MAE) is calculated by the energy difference between states with out-of-plane and in-plane magnetization, as plotted in Fig. \ref{fig4}(c). It is found that the MAE is strongly dependent on the magnetic phase in trilayer CrI$_3$, and in the strain range of -6\%$\sim$0\%, the stacking structures \Rmnum{1} to \Rmnum{4} possess the FM-$\uparrow\uparrow\uparrow$ state, and their MAE are almost the same and decrease with the increase of lattice constant. With the application of tensile strain, the magnetic phase of stacking structures \Rmnum{1} to \Rmnum{4} becomes different, and the values of their MAE are different.

The Curie temperatures T$_C$ for \Rmnum{2} (monoclinic) and \Rmnum{3} (hexagonal) stacking structures in trilayer CrI$_3$ with FM-$\downarrow\uparrow\downarrow$ state are simulated by Monte Carlo (MC) simulation based on a Hamiltonian described in Eqs. (\ref{H}) - (\ref{Hinter1}). A 40$\times$40$\times$1 supercell of hexagonal lattice with periodic boundary conditions is adopted, and the MC steps for each temperature is 10$^6$. T$_C$ as a function of applied strain is plotted in Fig. \ref{fig4}(d). According to our calculations, in both monolayer and trilayer CrI$_3$, the multiples of experimental T$_C$ and simulated T$_C$ are close to 0.58 (see Supplemental Materials). Thus, to avoid the problem of overestimating T$_C$ in theoretical studies and better compare with the experimental results, the T$_C$ value obtained from the MC simulation for trilayer CrI$_3$ in our calculations are all rescaled by a factor of 0.58 to reproduce the experimental result. It is noted that the rescaled T$_C$ increases with the increase of tensile stain for small strain, which is qualitatively consistent with the experimental observation of T$_C$ \cite{Song2019}.

\subsection{E. Effect of electronic correlation}
\textcolor{black}{For 3$d$ orbitals in transitions-metal compounds, the electronic correlation parameter $U$ is important, and the estimated value for Cr atom in CrI$_3$ via the linear response approach \cite{Cococcioni2005} is about 3.35 eV (see Supplemental Materials). Thus, $U$ = 3 eV is adopted in our above DFT calculations. To verify the robustness of our DFT results, we have also studied the magnetic phases of the four stacking structures under different strains with $U$ = 4 eV. It is shown that our conclusion does not change for $U$ = 3 and 4 eV.} As shown in Fig. S8 in Supplemental Materials, for all four stacking structures, by applying a small tensile strain, the FM-$\uparrow\uparrow\uparrow$ magnetic state remains for \Rmnum{1} (rhombohedral) stacking structure, while it changes to FM-$\downarrow\uparrow\downarrow$ for \Rmnum{2} (monoclinic) and \Rmnum{3} (hexagonal) stacking structures, and it transforms to FM-$\uparrow\uparrow\downarrow$ for \Rmnum{4} (triclinic) stacking structure. So, our conclusions do not change for reasonable parameter range of $U$ = 3 $\sim$ 4 eV. \textcolor{black}{In addition, for the DFT+$U$ scheme with $U$ ranging from 0 to 4 eV, the picture of S=3/2 with all spin up t$_{2g}$ orbitals occupied is reasonable (see Fig. S2 in Supplemental Materials). The crystal-field splitting between t$_{2g}$ and e$_g$ orbitals of the trilayer CrI$_3$ with \Rmnum{1} (rhombohedral) stacking order can be roughly estimated as 3.62 eV, which changes slightly with the strain (see Fig. S3 in Supplemental Materials).}

\section{\uppercase\expandafter{\romannumeral6}. CONCLUSION}
In summary, by using the first-principles calculations, we find that the magnetic states in trilayer CrI$_3$ strongly depend on the stacking structures and interlayer distance. All four different stacking structures, including \Rmnum{1} (rhombohedral), \Rmnum{2} (monoclinic), \Rmnum{3} (hexagonal) and \Rmnum{4} (triclinic), the FM-$\uparrow\uparrow\uparrow$ is the magnetic ground state. Under a small tensile strain, FM-$\uparrow\uparrow\uparrow$ is maintained in \Rmnum{1} (rhombohedral), while it transforms to FM-$\downarrow\uparrow\downarrow$ for \Rmnum{2} (monoclinic) and \Rmnum{3} (hexagonal), and changes to FM-$\uparrow\uparrow\downarrow$ for \Rmnum{4} (triclinic). The three obtained magnetic states are consistent with the observations in the recent experiment. The intralayer and interlayer exchange couplings are extracted based on a Heisenberg-type Hamiltonian, which can well interpret the change of magnetic behavior in trilayer CrI$_3$. Our results also show that the band gap and MAE of trilayer CrI$_3$ are strongly dependent on the magnetic phases. This study presents a reasonable explanation for the experimental observations for the trilayer CrI$_3$, and paves a way to design multilayer devices with desired properties, such as novel magnetic states, suitable band gaps, and so on.

\section{Acknowledgements}

This work is supported in part by the National Key R$\&$D Program of China (Grant No. 2018YFA0305800), the Strategic Priority Research Program of the Chinese Academy of Sciences (Grant No. XDB28000000), the National Natural Science Foundation of China (Grant No.11834014), and Beijing Municipal Science and Technology Commission (Grant No. Z191100007219013). B.G. is also supported by the National Natural Science Foundation of China (Grants No. Y81Z01A1A9 and No. 12074378), the Chinese Academy of Sciences (Grants \textcolor{black}{No. YSBR-030}, No. Y929013EA2 and No. E0EG4301X2), the University of Chinese Academy of Sciences (Grant No. 110200M208), the Strategic Priority Research Program of Chinese Academy of Sciences (Grant No. XDB33000000), and the Beijing Natural Science Foundation (Grant No. Z190011).


%

\end{document}